# Current-Driven Dynamics of Skyrmions Stabilized in MnSi Nanowires Revealed by Topological Hall Effect

*Dong Liang*[1†], *John P. DeGrave*[1†], *Matthew J. Stolt*[1], *Yoshinori Tokura*[2], *Song Jin*[1]*

[1] Department of Chemistry, University of Wisconsin-Madison, Madison, Wisconsin 63706, United States

[2] RIKEN Center for Emergent Matter Science (CEMS), Wako 351-0198 and

Department of Applied Physics and Quantum-Phase Electronics Center (QPEC), University of Tokyo, Tokyo 113-8656, Japan

† These authors contributed equally to this work.

*email: jin@chem.wisc.edu

**Abstract:**

Skyrmions, novel topologically stable spin vortices, hold promise for next-generation magnetic storage due to their nanoscale domains to enable high information storage density and their low threshold for current-driven motion to enable ultralow energy consumption. One-dimensional (1D) nanowires are ideal hosts for skyrmions since they not only serve as a natural platform for magnetic racetrack memory devices but also can potentially stabilize skyrmions. Here we use the topological Hall effect (THE) to study the phase stability and current-driven dynamics of the skyrmions in MnSi nanowires. The THE was observed in an extended magnetic field-temperature window (15 to 30 K), suggesting stabilization of skyrmion phase in nanowires compared with the bulk (27 to 29.5 K). Furthermore, for the first time, we study skyrmion dynamics in this extended skyrmion phase region and found that under the high current-density of $10^8$-$10^9$ Am$^{-2}$ enabled by nanowire geometry, the THE decreases with increasing current densities, which demonstrates the current-driven motion of skyrmions generating the emergent electric field. These results open up the exploration of nanowires as an attractive platform for investigating skyrmion physics in 1D systems and exploiting skyrmions in magnetic storage concepts.

A magnetic skyrmion is a vortex-like spin texture in which the central spin is anti-aligned with the externally applied magnetic field and the outer spins are aligned[1,2]. Their formation results from the competition of the ferromagnetic exchange and the Dzyaloshinskii-Moriya interaction that appears in crystal systems lacking a center of inversion symmetry, such as the non-centrosymmetric cubic B20 crystal system, including $MnSi$[3-5], $Fe_{1-x}Co_xSi$[6], $FeGe$[7], $MnGe$[8]. Skyrmion domains can be manipulated using remarkably low current densities compared to ferromagnetic domain walls, which has become the focus of intense interest for future low-power consumption nanoscale spintronics devices[2,9-12]. The relative stability and ease of manipulating magnetic skyrmions with the electrical current result from their non-trivial topology, smoothly varying spin configuration, and unique ability to deform and avoid pinning sites[2,13]. This ultra-low current density may enable low-power consumption applications, and help to avoid common failure modes associated with very large current densities encountered in ferromagnetic systems being explored for magnetic racetrack memory concepts[14].

To explore skyrmions for potential magnetic storage technology, we need to study the stability and dynamics of skyrmions in skyrmion-hosting nanoscale systems[12,15,16]. In particular, one-dimensional (1D) nanowires (NWs) not only serve as a natural platform for magnetic racetrack memory devices[14] using skyrmions as information units, but also could have significant advantages over bulk materials to stabilize skyrmions[17,18]. For example, as the dimension of the system approaches the size of the helical pitch that characterizes the helimagnetic ground state, the conical magnetic configuration will be destabilized energetically compared to the skyrmion state[17]. This has been experimentally observed in thin films of B20 materials with nanoscale thickness[5,7,19,20]. The stabilization of the skyrmions was also theoretically predicted in nanoscale systems by the introduction of large anisotropy energy[18]. Furthermore, anisotropic 1D NWs with small cross-sections are natural platforms for studying emergent electrodynamics associated with current-driven skyrmions at much larger current densities than can ever be achieved in bulk. Therefore, NW systems can be utilized to manipulate skyrmions on faster timescales and study the fundamental dynamics of the skyrmions[2,15,16].

Skyrmions and their current-driven motion in B20 crystals have been studied by *k*-space observation using small-angle neutron scattering in bulk crystals of MnSi[9] and by real space observation using Lorentz transmission electron microscopy (LTEM) in FeGe thin films[10]. These experiments have shown that the skyrmion domains translate primarily along the current direction at a current density on the order of $10^6$ A m$^{-2}$, which is up to 4-5 orders of magnitude smaller than the current densities required for the translation of ferromagnetic and helimagnetic domain walls[13,21]. Even though stabilization of the skyrmion lattice in MnSi NWs has been suggested using LTEM observations of a thinned NW slab[5], magnetoresistance[22], and most recently dynamic cantilever magnetometry study[23], the dynamics of skyrmions in NW systems is more difficult to investigate. Hall transport measurements, especially the observations of topological Hall effect (THE) beyond the normal Hall effect (NHE) and anomalous Hall effect (AHE), have been considered as an electrical transport signature of skyrmions[2,11,19,20,24,25]. This is because when a conduction electron passes through a topologically non-trivial spin texture, the spin of the conduction electron adiabatically couples to the local spin and acquires a quantum-mechanical Berry phase that can be re-formulated in terms of an effective magnetic field[2,11], which deflects the conduction electrons perpendicular to the current direction. Therefore, the presence of skyrmions will cause an additional contribution to the observed Hall signal which has been termed THE. Furthermore, when the skyrmions start to move above the critical current density, the THE effect is predicted[26] and observed[11] to decrease due to suppression of the THE by the emergent electric field. If the skyrmion domains are to be exploited for magnetic storage applications, emergent electrodynamics study of the current-driven skyrmion motion need to be extended to nanoscale geometries which allow much higher current densities and potentially higher velocities of the skyrmion motion to be studied[13]. Therefore, despite the challenging 1D NW geometry for fabricating Hall devices[27], there exists a strong motivation to investigate skyrmion dymanics of NWs using the THE.

Here, using Hall effect measurements of MnSi NWs, we report the observation of the topological Hall effect due to the stabilized skyrmions and the study of their current-driven dynamics at large current densities ($10^8 – 10^9$ A m$^{-2}$). We clearly demonstrate the extraction of the topological portion of the Hall signal from the total Hall signal and use the THE to construct magnetic field ($B$)-temperature ($T$) phase diagrams for MnSi NWs, which show that skyrmions are stable in these 1D MnSi NWs over a larger $B$-$T$ range. Furthermore, for the first time, we study the electrodynamics of current-driven skyrmions in NW morphology at large current densities and show the suppression of the THE due to the emergent electric field arising from current-driven motion of skyrmions and estimate the skyrmion drift velocity.

**Results and Discussion**

The Hall devices were fabricated using single crystal MnSi NWs synthesized by chemical vapor deposition[28] and using an improved device fabrication strategy, in which three-step angled evaporations of metal electrodes (see Fig. S1 in the Supplementary Information) ensure side-wall

Hall electrode contact[29], but now the opposing Hall electrodes are aligned as in the classical Hall bar devices to maximize the Hall signal. The inset of Fig. 1a shows a typical SEM image of a representative MnSi NW Hall device. Both transverse (Hall) and longitudinal resistance in magnetic field were measured. We will focus the discussion on the data from two representative MnSi NW devices, namely NW1 (330 nm wide by 270 nm thick) and NW2 (shown in Fig. 1a inset, 480 nm wide by 300 nm thick). We only collected data in the magnetic field range from −1 T to +1 T, which was determined to be the region of the greatest interest through preliminary scans for identifying the THE attributed to the skyrmion phase in MnSi NWs, and was additionally suggested by previous Lorentz TEM observations of a MnSi NW from the same synthesis[5]. Fig. 1a show the raw transverse resistance measured between opposing Hall-bar electrodes of NW1[29] at $T = 22$ K and current $I = 80$ µA. The data were obtained by sweeping magnetic field in both positive (-1 to 1 T) and negative (1 to -1 T) directions as indicated by blue and red dots, respectively. The two curves show no significant hysteresis and overlap very well with a very tiny deviation, which is consistent with the behaviors of helical and conical phase. Then we took average of the two curves and antisymmetrized the averaged data to obtain actual Hall resistance ($R_{yx}$). After that, the Hall resistivity ($\rho_{yx}$), shown as the red curve in Fig. 1b, is calculated by

$$\rho_{yx} = R_{yx} t \qquad (1)$$

where $t$ is the thickness of the NW.

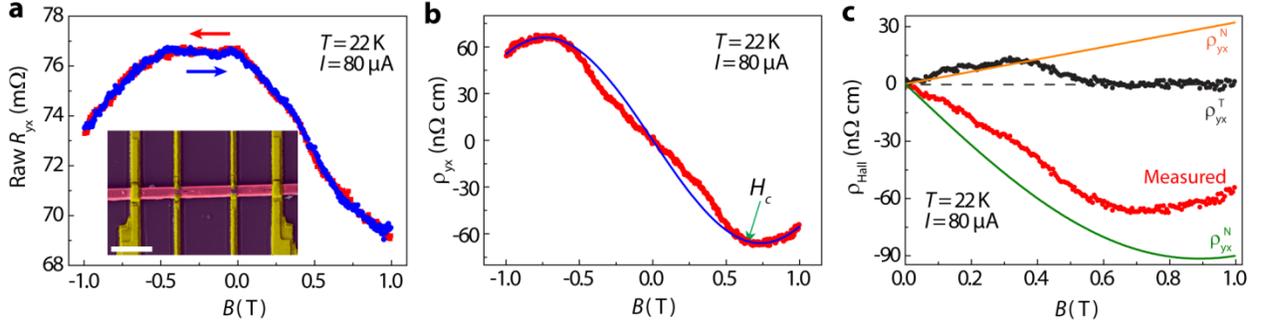

**Figure 1 | Extraction of the topological Hall effect resistivity ($\rho_{yx}^T$) with external magnetic field perpendicular to the NW1 at $T$ = 22 K and $I$ = 80 μA (current density $j = 9.16 \times 10^8$ Am$^{-2}$). a,** The raw Hall resistances shown as red (positive field sweep from -1 to 1 T) and blue (negative field sweep from 1 to -1 T) curves measured between the transverse Hall electrodes. Inset shows a typical NW device with classical Hall bar. The scale bar is 2 μm. **b,** The antisymmetric Hall resistivity (ρ$_{yx}$) extracted from the raw Hall resistance in panel a. The blue line shows the polynomial fitting above critical field $H_c$ and extrapolation below $H_c$ in order to simulate the normal and anomalous Hall contributions to the total Hall resistivity. **c,** Extracted topological Hall signal in comparison with normal and anomalous Hall contributions. The topological Hall resistivity ($\rho_{yx}^T$) in black curve is extracted by subtracting the fitting (blue curve in panel b) from the total measured Hall signal (red curve in panel b). The fitting curve is also decomposed to normal (orange line) and anomalous (green curve) Hall contributions.

We will use the representative data shown in Fig. 1b to illustrate the procedures for extracting the THE. The total Hall resistivity (ρ$_{yx}$) measured in the MnSi NW in the region of the skyrmion phase consists of three components of resistivity,

$$\rho_{yx} = \rho_{yx}^N + \rho_{yx}^A + \rho_{yx}^T \tag{2}$$

Where the three terms on the right side correspond to the NHE attributed to Lorentz forces on the conduction electrons due to the externally applied magnetic field, the AHE which is known to scale with the magnetization of the sample, and the THE which is attributed to the emergent magnetic fields associated with the skyrmion domains, respectively. Above the critical field ($H_c$) for the transition between the conical and the field-induced ferromagnetic state, the dominant contribution to the measured Hall signal is the NHE and AHE. The NHE is expressed as $\rho_{yx}^N = R_0 B$, where $R_0$ is the normal Hall coefficient. The AHE contribution ($\rho_{yx}^A$) is expressed as $\rho_{yx}^A = \alpha M \rho_{xx0} + \beta M \rho_{xx0}^2 + b M \rho_{xx}^2$ where $\alpha$, $\beta$, and b correspond to the skew scattering, side jump, and intrinsic contributions to the anomalous Hall resistivity, $\rho_{xx0}$ is the residual resistivity, and $M$ is the magnetization[20,30]. The THE can be extracted by subtracting the NHE and AHE from total Hall signal. Although recent cantilever magnetometry on a single NW allowed measurement of $M$ in a parallel geometry ($H$ // NW long axis)[23], the direct measurement of $M$ in NWs in perpendicular $B$ is still convoluted. Instead we must rely on a phenomenological assumption for the AHE inferred from previous works on bulk MnSi[24] and thin films of MnSi[20]. Namely, we assume that the magnetization is smoothly varying in the region of skyrmion phase, together with the fact that AHE and NHE dominate above the critical field ($H_c$), the AHE and NHE can be simulated by fitting the total Hall effect signal above the critical field ($H_c$) and extrapolating below $H_c$, as shown in the blue curve in Fig. 1b. We accomplish this fitting by using a degree-five polynomial function $aB + bB^3 + cB^5$ with fitting parameters of $a, b, c$ to represent the NHE and AHE contributions in the total magnetic field of interest. Then we perform the subtraction of simulated NHE and AHE to obtain THE confidently in the suspected region of the skyrmion formation based both upon previous transport studies of MnSi[24,31] as well as recent Lorentz-TEM results of a thinned MnSi NW specimen[5]. We note that in principle a similar phenomenological data extraction was applied to recent experiments of the topological

Hall effect anomaly in bulk MnSi in which the non-topological Hall signal was fit using the assumption of a linear background[11,32]. The extracted THE signal is shown in Fig. 1c as the black curve. The blue fitting curve in Fig. 1b is decomposed to NHE (orange line) and AHE (green curve) components in Fig. 1c. We notice the NHE at all temperatures of interest indicates hole-like carriers, which is consistent with those reported for MnSi bulk[24,31] and thin film[20]. In contrast, the AHE is negative and dominates the Hall signal. The extracted THE has the same sign as the NHE, but opposite sign to the AHE. These observations are consistent with those reported in MnSi thin film[20].

Hall resistivities at various temperatures from 10 to 34 K are plotted in Fig. 2a. We then extracted the topological Hall resistivity ($\rho_{yx}^T$) via subtracting the polynomial fitting from $\rho_{yx}$ at various temperatures, as discussed above. The extracted $\rho_{yx}^T$ values for NW1 from 10 to 34 K with the step of 2 K are shown in Fig. 2b. The magnitude of the THE ($\rho_{yx}^T$) is ~15 nΩ cm at 22 K, slightly larger than that measured in MnSi bulk[11,24] and thin film[20]. We further use this set of data to construct the interpolated pseudo-color *B-T* phase diagram shown in Fig. 2c. To show the generality of the topological Hall signal, we performed the same analysis on a second MnSi NW device (NW2) and constructed the phase diagram for the $\rho_{yx}^T$ shown in Fig. 2d. For both NW1 and NW2, the skyrmion phase has been identified in an extended phase region from a minimum of 15 K up to a $T_c$ of roughly 30 K and a magnetic field region from 0.1 to 0.5 T. Note that the temperatures reported in the phase diagrams (Fig. 2c, d) can be slight underestimates of the true NW device temperatures because of the Joule heating effect, which will be discussed in detail later.

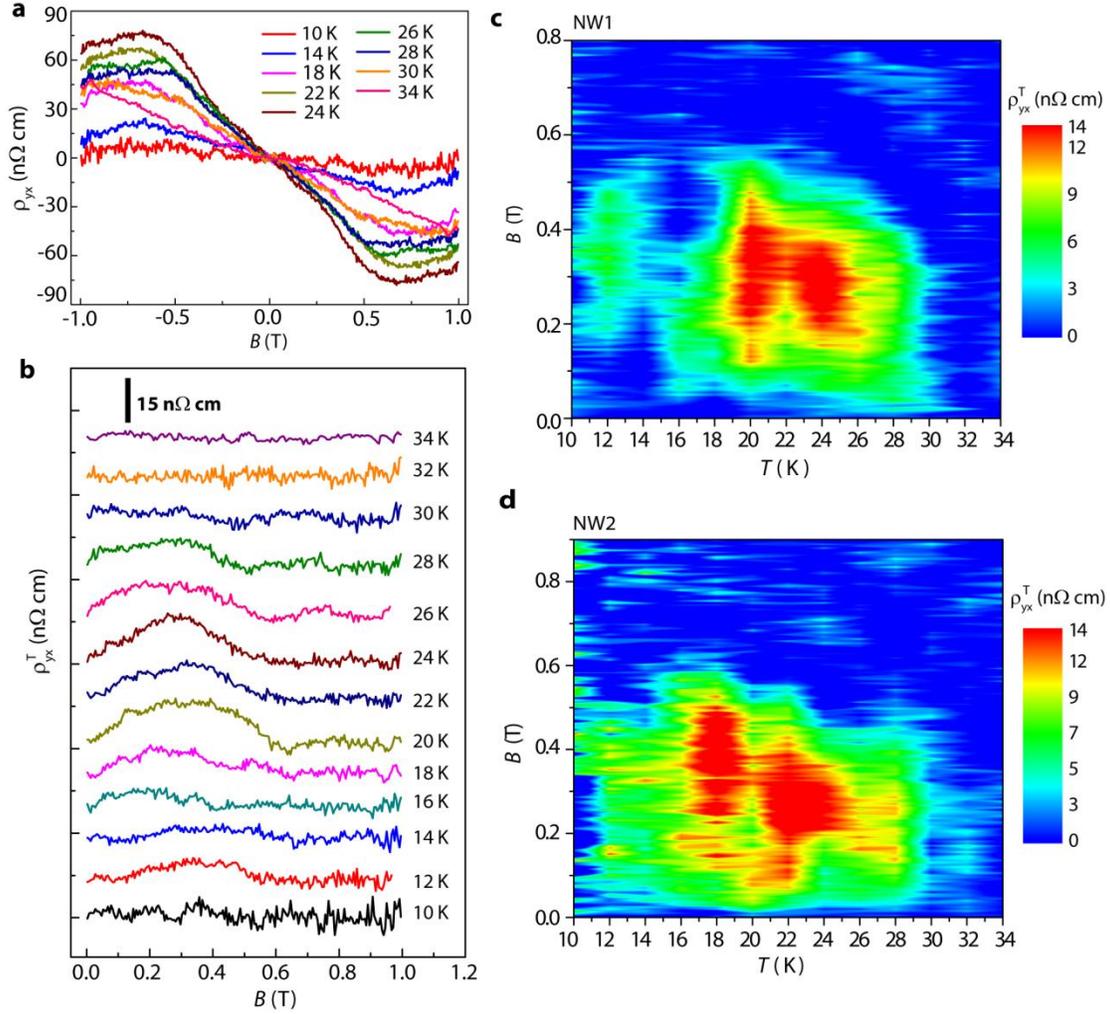

**Figure 2 | Topological Hall resistivity and the magnetic phase diagrams for two MnSi NW devices. a,** Hall resistivity from anti-symmetrization of raw data from 10 to 34 K and **b,** Topological Hall resistivity ($\rho_{yx}^T$) for NW Device 1 at several temperatures extracted using the polynomial fitting procedure detailed in the text. **c** and **d,** Phase diagrams of extracted $\rho_{yx}^T$ as a function of the external magnetic field (*B*) and temperature (*T*) for NW1 and NW2 constructed by interpolating $\rho_{yx}^T$ between temperatures. All data are collected at *I* = 80 μA, corresponding to a current density of $9.16 \times 10^8$ Am$^{-2}$ for NW1 and $5.52 \times 10^8$ Am$^{-2}$ for NW2.

The magnitude of $\rho_{yx}^T$ is known to depend on several factors and can be written as $\rho_{yx}^T \approx P \frac{q_\sigma^e}{e} R_0 B^e$ [24], where $P$ is the spin polarization of charge carriers, $q_\sigma^e$ is the emergent charge associated with the charge carrier spins ($q_\uparrow^e = +1/2$ and $q_\downarrow^e = -1/2$), $e$ is the elementary charge, $R_0$ is the normal Hall coefficient, and $B^e$ is the emergent magnetic field associated with one unit cell of the skyrmion lattice. $B^e$ is proportional to the skyrmion winding number $\phi = \frac{1}{4\pi} \vec{n} \cdot \frac{\partial \vec{n}}{\partial x} \times \frac{\partial \vec{n}}{\partial y}$ where $\vec{n} = \frac{\vec{M}_s}{|M_s|}$ is the local magnetization direction, and $x$ and $y$ are the coordinates perpendicular to the applied magnetic field[2]. The value of $B^e$ is roughly 13.1 T$|e/q^e|$ for the hexagonal lattice of skyrmions in MnSi[24]. Here for MnSi NWs, Lorentz TEM study[5] showed no change to the size and density of skyrmions, so we don't expect much change in $B^e$. However, $R_0$ in our NWs is about twice as large as that in bulk MnSi and the spin polarization ($P$) might also change in NWs, therefore it is not unexpected that $\rho_{yx}^T$ of magnitude ~15 nΩ cm is higher than those in bulk[11,24] and thin film[20]. Note that the observed $\rho_{yx}^T$ = 15 nΩ cm at 80 µA is not the THE for stationary skyrmions, later it will be shown that at 80 µA skyrmions are in motion.

The temperature window in which we observe the THE signal in MnSi NWs is roughly between 15 to 30 K, which places these MnSi NW samples somewhere between the observed behavior in bulk MnSi with a narrow skyrmion phase region (27 – 29.5 K)[24], and the observed behavior for strained 50 nm thick MnSi thin films grown by molecular beam epitaxy with a wider skyrmion phase region (2 – 45 K)[20]. Since the width and thickness of our NW samples on the order of 300 nm are much larger than the skyrmion size of 18 nm in MnSi, spatial confinement effect on skyrmions can be excluded. Uniaxial magnetic anisotropy, from intrinsic strain or surface effect, is more likely to induce the extension of the THE region[18]. In particular, large surface-to-volume ratio of NWs is likely to manifest the surface effect-induced uniaxial anisotropy and hence could explain the extended stabilization of skyrmions in MnSi NWs[5,23]. We shall clarify that the THE measurements detect the spin chirality in the system, so the extracted skyrmion phase above can either be skyrmion lattice or possibly be mixture of skyrmions with other magnetic states, as also suggested by cantilever magnetometry[23].

The magnetic anisotropy in MnSi NWs is further supported by the influence of external magnetic field orientation on the skyrmion phase stability inferred from the THE in perpendicular magnetic field and from the longitudinal magnetoresistivity in parallel field. For the latter measurements, the NW long axis was aligned under an optical microscope within a 10° error before mounting the device chip. This magnetic field orientation allows us to observe the effects of the strong anisotropy from the NW morphology on the formation of the skyrmion phase. Similar to what we have reported[22], the longitudinal magnetoresistivity of NW1 (Fig. 3a) shows three kinks (highlighted by dash lines) that are clear signatures of three critical fields identified as: (1) $H_{//c}$ (commonly referred to as $B_{C2}$ in MnSi literature) corresponding to the transition field from conical to field-polarized ferromagnetic states; (2) $B_{//A2}$ corresponding to the transition field from skyrmion to conical states; and (3) $B_{//A1}$ as the transition field from helical to skyrmion states. These critical fields have been identified in magnetoresistance of MnSi bulk[33] and NW samples[22] and have been attributed to the skyrmion phase. The $B_{//A1}$ (filled circles), $B_{//A2}$ (diamonds), $H_{//c}$ (stars) determined by the kinks in magnetoresistivity of NW1 in a parallel magnetic field are plotted in Fig. 3b. The skyrmion phase indicated by longitudinal magnetoresistivity persists from 30 K down to the lowest measurement temperature 10 K in parallel magnetic field orientation (actually it can go down to 3 K according to our previous

report[22]), whereas the skyrmion phase inferred from the THE measurements exist from 15 K to 30 K in perpendicular orientation. These differences in skyrmion phase diagrams between the two orientations indicate strong magnetic anisotropy that could enable a positive uniaxial anisotropy to stabilize the THE (and skyrmion) phase, since theory predicts that the uniaxial anisotropy in B20 crystals is essential in stabilization of skyrmions[18,19].

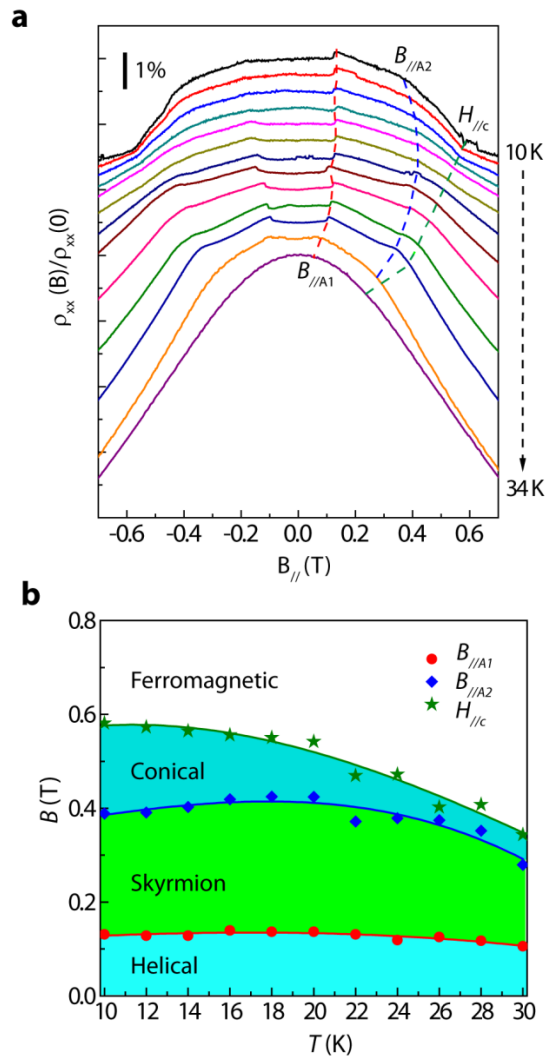

**Figure 3 | Magnetoresistivity of NW1 measured at $I$ = 80 μA** (current density $j = 9.16 \times 10^8$ Am$^{-2}$) **with the externally applied magnetic field preferentially parallel to the long axis of**

the NW (<10 °misalignment) and the corresponding magnetic phase diagram. a,** Normalized longitudinal magnetoresistivity measured from 10 to 34 K with a step of 2 K. The dash lines indicating the approximate positions of critical fields corresponding to the $B_{//A1}$ (red filled circles), $B_{//A2}$ (blue diamonds) and $H_{//c}$ (green stars) in the *B-T* phase diagram in panel b. **b,** Magnetic phase diagram obtained from the longitudinal magnetoresistivity in a parallel magnetic field. The phase boundaries revealed by longitudinal magnetoresistivity are represented by red filled circles ($B_{//A1}$), blue diamonds ($B_{//A2}$) and green stars ($H_{//c}$), which correspond to transitions from helical to skyrmion states, from skyrmion states to conical states, and from conical to field-polarized ferromagnetic states, respectively. Skyrmion phase is shown to persist from 30 K down to 10 K (the lowest temperature in our measurements).

Furthermore, we studied the current-density dependence of the THE in MnSi NWs. We typically used applied currents of between 40 and 240 μA, which corresponds to the current density regime of $10^8 – 10^9$ Am$^{-2}$, depending on the NW cross section dimensions. The maximum current densities applied to NW1 and NW2 were on the order of $10^9$ A m$^{-2}$ ($2.75 \times 10^9$ A m$^{-2}$ for NW1 and $1.38 \times 10^9$ A m$^{-2}$ for NW2), which are roughly three orders of magnitude larger than the critical current density required for depinning the skyrmion lattice observed for high purity bulk crystals of MnSi, yet still much lower than the upper threshold current density (on the order of $10^{12}$ A m$^{-2}$) beyond which the melting of the skyrmion lattice into a chiral liquid phase is suggested to occur by a recent calculation[34]. Such high current densities enabled by the NW geometry of our samples offer a unique regime to study the skyrmion dynamics.

The side-effect of such high current densities is the local Joule heating of the NW device. For currents larger than 40 μA, both devices showed non-ohmic *I-V* curves (or increase of resistance) and decrease of critical field $H_c$, as shown in Fig. S2. This indicates large currents raised the local device temperature to a value higher than the nominal cryostat temperature. Significant Joule heating effect, observed at currents larger than 80 μA, made it difficult to examine the current-density dependence of the THE at various current densities while keeping the same temperature. In order to overcome this hurdle, we applied additional cryostat cooling to compensate for the temperature increase due to Joule heating. Initially, at 20 μA with negligible Joule heating, the actual device temperature was determined by the nominal cryostat temperature. The four-probe resistance of the device at 20 μA served as the temperature standard. At higher currents, Joule heating effect led to the increase of local temperature and hence the resistance. We lowered the cryostat nominal temperature to cool down the device and decrease the device resistance to the standard value at 20 μA. Following such strategies, we paired current values with nominal cryostat temperatures one by one so that temperatures at higher currents are calibrated to the same value at 20 μA (see details in Supplementary Information). Therefore, THE measurements are accessible at the same temperature but different currents in order to investigate the current-density dependence of the THE. Fig. S3 in Supplementary Information shows the temperature compensation chart in which pairs of applied currents and nominal cryostat temperatures correspond to the same resistance and hence the same temperature.

Fig. 4a presents topological Hall resistivity at various current densities ($j$) of device NW1 at $T = 24.04$ and $22.12$ K. The average topological Hall resistivity from 0.2 to 0.4 T vs. current density ($j$) is shown in Fig. 4b. The similar set of data for device NW2 are presented in Fig. 4d-e. It is clear that the THE decreases with increasing current densities from $10^8$ A m$^{-2}$ to $10^9$ A m$^{-2}$ and then $\rho_{yx}^T$ appears to show a plateau below about 10 nΩ cm. These observations are consistent with the movement of skyrmion domains driven by electrical current and the theoretical prediction that the emergent electric field induced by motion of skyrmions suppresses the THE, as we will discuss in detail below.

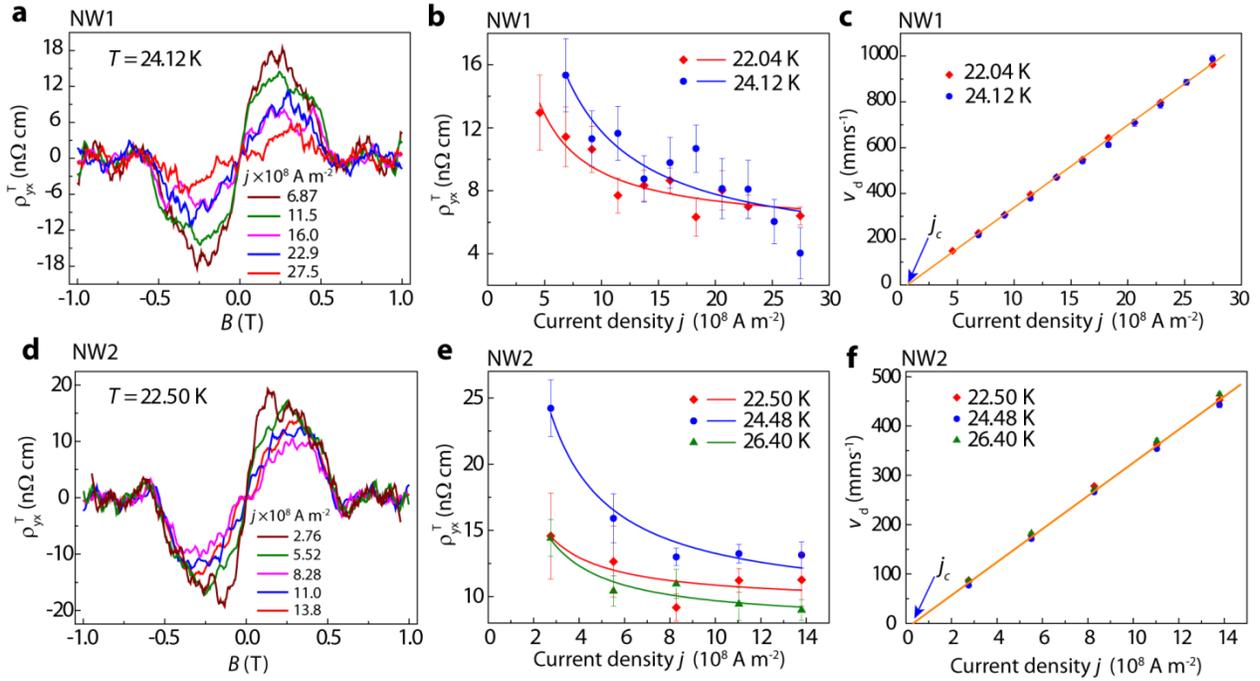

**Figure 4 | Topological Hall resistivity ($\rho_{yx}^{T}$) of MnSi NWs as a function of current density (*j*). a,** $\rho_{yx}^{T}$ *vs. B* at $T = 24.12$ K and **b,** $\rho_{yx}^{T}$ *vs. j* at $T = 22.04$ and $24.12$ K under increasing current densities from $j = 6.87 \times 10^8$ to $27.5 \times 10^8$ A m$^{-2}$ for device NW1. Solids lines represent nonlinear fitting based on equation (4). **c,** Estimated skyrmion drift velocity ($v_d$) as a function of current density (*j*). **d,** $\rho_{yx}^{T}$ *vs. B* at $T = 22.50$ K and **e,** $\rho_{yx}^{T}$ *vs. j* at $T = 22.50, 24.48, 26.40$ K under increasing current densities from $j = 2.76 \times 10^8$ to $13.8 \times 10^8$ A m$^{-2}$ for device NW2. Solids lines represent nonlinear fitting based on equation (4). **f,** Estimated skyrmion drift velocity ($v_d$) as a function of current density(*j*).

To explain the observed current density-dependent THE, we will begin by providing a brief introduction to skyrmion dynamics. The spin-polarized current exerts a force on the underlying magnetic spin structure through the spin-transfer-torque mechanism, a phenomenon that has been explored extensively in ferromagnetic systems[13,35,36]. At sufficiently high current densities this force is large enough to overcome the pinning forces of the skyrmion domains, and the skyrmion domains begin to translate, as has been observed by the neutron diffraction in bulk MnSi crystals[9] and Lorentz TEM in FeGe thin films[10]. In analogy with Faraday's law, the motion of the effective magnetic field is expected to cause an "emergent" electric field perpendicular to the motion of the skyrmions $\vec{E}^e = -P\frac{q^e_\sigma}{e}\vec{v}_d \times \vec{B}^e$ where $\vec{B}^e$ is the emergent magnetic field associated with a skyrmion, $\vec{v}_d$ is the drift velocity of skyrmions. This emergent electric field opposes the topological Hall field arising from the Berry phase of the stationary skyrmions[26,2], leading to a reduction of the measured $\rho^T_{yx}$. So the net transverse topological electrical field $E^T_y = P\left|\frac{q^e}{e}\right|(v-v_d)B^e$ depends on the relative drift velocity between skyrmions and charge carriers, where $v$ is charge carrier drift velocity. Then the current density dependent topological Hall resistivity can be expressed as[11]

$$\rho^T_{yx}(j) = \frac{E^T_y}{j_x} = \frac{P\left|\frac{q^e}{e}\right|(v-v_d)B^e}{nev} = P\left|\frac{q^e}{e}\right|R_0 B^e\left(1-\frac{v_d}{v}\right) \quad (3)$$

Above the critical current density $j_c$ (and the critical charge carrier drift velocity $v_c$), the skyrmion drift velocity ($v_d$) is linearly proportional to carrier drift velocity ($v$), as suggested by the theory[13,26] and confirmed by experiment[11]. Here we can define a relation $v_d = A(v-v_c)$ by introducing coefficient $A$. Together with $v = jR_0$, equation (3) turns into

$$\rho^T_{yx}(j) = P\left|\frac{q^e}{e}\right|R_0 B^e\left(1-A\frac{(v-v_c)}{v}\right) = \rho^T_{yx}(j \leq j_c)\left(1-A\left(1-\frac{j_c}{j}\right)\right) \quad (4)$$

Equation (4) shows how the topological Hall resistivity evolves with the applied current density. At or below $j_c$, $\rho_{yx}^T(j \leq j_c) = P\left|\frac{q^e}{e}\right|R_0 B^e$ is the topological Hall resistivity for stationary skyrmions. Slightly or moderately above $j_c$ where skyrmions start to move, $\rho_{yx}^T(j > j_c) \propto 1/j$. Far above $j_c$ where $1/j$ approaches zero, $\rho_{yx}^T(j \gg j_c) \approx P\left|\frac{q^e}{e}\right|R_0 B^e (1-A)$, which indicates $\rho_{yx}^T(j)$ saturates to the value with a factor $(1-A)$ lower than that for stationary skyrmions, which is also predicted by Ref. 26. Equation (4) also provides the guideline for searching for skyrmion dynamics via observation of the current density dependent THE. The appropriate observation window should start right around the critical current density $j_c$ and should not exceed the current density at which the topological Hall resistivity starts to saturate.

The data in Fig. 4b and 4e agree well with the relation $\rho_{yx}^T(j > j_c) \propto 1/j$, indicating the dynamics of skyrmions in MnSi NWs are in a current regime moderately above $j_c$. We can estimate the parameters in equation (4) in MnSi NWs as following: current polarization $P$ is estimated to be 0.2 [32], emergent magnetic field is taken as $B^e$ = 13.1 T$|e/q^e|$, and Normal Hall coefficients were measured to be $R_0$ = 32.5 and 37.1 nΩ cm/T for NW1 and NW2, respectively. Therefore $\rho_{yx}^T(j \leq j_c)$ is estimated to be 85.2 and 97.2 nΩ cm for NW1 and NW2, respectively. Using equation (4) to fit the $\rho_{yx}^T(j)$ vs. $j$ data (Fig. 4b, 4e) can then allow us to extract fitting parameters $A$ and $j_c$. Coefficients ($A$) and critical currents ($j_c$) are obtained and averaged across different temperatures to yield $A$ = 0.95, $j_c$ = 7.2×10$^7$ A m$^{-2}$ for NW1, and $A$ = 0.91, $j_c$ = 2.7×10$^7$ A m$^{-2}$ for NW2. These critical current densities on the order of 10$^7$ A m$^{-2}$ are higher than the ~10$^6$ A m$^{-2}$ observed in MnSi bulk. The possible explanation could be that skyrmions tend to be deflected to one sidewall of the NW due to skyrmion Hall effect[2] so that skyrmions are more easily exposed to surface defects in the NW with high aspect-ratio, which could result in the increase of pinning force of skyrmions.

We then calculate and fit the skyrmion drift velocity ($v_d$) vs. $j$ for NW1 (Fig. 4c) and NW2 (Fig. 4f) using:

$$v_d = jR_0\left(1 - \rho_{yx}^T(j)/\rho_{yx}^T(j \leq j_c)\right) \quad (6)$$

At $j \sim 10^9$ A m$^{-2}$, $v_d$ is ~ 330 mms$^{-1}$ for both NW1 and NW2. For each NW device, the $v_d$ for all temperatures follows one common linear dependence on $j$ as indicated by the orange solid lines in Fig. 4c and 4f. The slopes of these lines are proportional to coefficients $A = \Delta v_d / \Delta v$ (0.95 for NW1 and 0.91 for NW2). This means that the skyrmion drift velocity ($v_d$) increases at a rate slower than the carrier drift velocity ($v$) with a factor of $A$. The behavior of skyrmion drift velocity is in accord with spin-transfer-torque mechanism[13,26]. Our results here uniquely show the skyrmion dynamics arising from the emergent electric field due to the motion of current-driven skyrmions at relatively large current densities in an extended skyrmion stability region and its interplay with the topological Hall effect in 1D NWs.

**Conclusion**

In summary, we have measured the topological Hall effect in MnSi NW devices under large current densities, which first shows that the inferred skyrmion phase is stabilized over an extended magnetic field ($B$)-temperature ($T$) window in MnSi NWs compared to bulk MnSi crystals. Furthermore, over a wider temperature window than for bulk materials, we show THE decreases with increasing current-density, indicating that the skyrmions are moving due to the electrical current and the THE is suppressed by the emergent electric field arising from motion of skyrmions. The extended skyrmion stability in MnSi NWs and the corresponding current-driven skyrmion dynamics at large current densities are interesting for the fundamental study of the skyrmion physics in confined and/or anisotropic systems, and more importantly demonstrate the feasibility and advantages of exploiting NW systems for potentially using skyrmions in magnetic memory storage systems.

**Methods**

Electrical measurements were implemented on MnSi NWs obtained via a chemical vapor deposition synthesis[28]. Classical Hall bars were fabricated on MnSi NWs using Ti/Au electrodes defined by e-beam lithography for ohmic electrical contacts. A three-step electrode deposition was employed as illustrated in Fig. S1, 20 nm Ti and 20 nm Au was evaporated vertically (the device chip plane is in 90 ° to evaporation direction) at first, followed by 60 nm Ti and 30 nm Au at a tilted angle of 45 ° exposing one side wall of NWs. Then the device chip was remounted and 90 nm Ti and 40 nm Au was evaporated at a tilted angle of 45 ° exposing the opposite side wall of NWs. Hall measurements were made in which the outermost electrical contacts act as current-source and drain, and the inner two contacts contacting opposite sides of the facetted MnSi NWs allow detection of the transverse Hall voltage (see Fig. 1a inset for a SEM image of a typical MnSi NW Hall device). The devices were measured in a Quantum Design PPMS-9T to provide the temperature and magnetic field control, with the DC current supplied by a Keithley 6221 current source and the voltage across the transverse contacts monitored by a Keithley 2182A nanovoltmeter.

The MnSi NWs were tested at room temperature for ohmic contact and were cooled from 300 K down to 10 K at a rate of 5 K/min under zero applied magnetic field. Once base temperature was reached, the device was allowed to equilibrate at 10 K for 30 min, and the field was set to -1 T. After collecting data across the field range from −1 T to +1 T the data collection was repeated in the reverse direction (from +1 T to −1 T) to check the reproducibility of the observed signals as well as to detect any magnetic hysteresis. After completion of the data collection to the final field of the second scan, the field was held at the final value of −1 T and the temperature was set to the next value at a rate of 2 K/min, and the device was allowed to equilibrate at the new set temperature for 20 min prior to data collection. The 20 min equilibration time was found to be more than sufficient to reduce any artifacts caused by thermal equilibration to a negligible level.

## Acknowledgements

This work is supported by US National Science Foundation (ECCS-1231916). Y.T. thanks the support by the Funding Program for World-Leading Innovative R&D on Science and Technology on "Quantum Science on Strong Correlation" and by JSPS Grant-in-Aid for Scientific Research No. 24224009. J.P.D. also thanks NSF and JSPS for jointly funding a fellowship partially supporting this collaboration through the EAPSI program. D.L. thanks Dr. Jiadong Zang and Dr. Yufan Li for helpful discussions.

## Author contributions

S.J., J.P.D., and D.L. conceived the project. D.L., J.P.D. carried out the device fabrication, electrical transport measurement, data analysis and theoretical modeling with M.J.S.'s assistance. Y.T. contributed to the data analysis and theoretical modeling. D.L., J.P.D. and S.J. wrote the manuscript with all authors contributing to the discussion and preparation of the manuscript.

**Conflict of interest**: the authors declare no conflicts of interest.